\documentclass[aps,pre,amsmath,superscriptaddress,twocolumn,longbibliography]{revtex4-1}

\usepackage[T1]{fontenc}
\usepackage{amsmath,amsfonts,amssymb,amsthm,bbm,bm, braket}
\usepackage{wasysym}
\usepackage{comment}
\usepackage{scalerel}
\usepackage{enumerate}
\usepackage{graphicx}
\usepackage{mathtools}
\usepackage{ifthen}
\usepackage{tensor}

\usepackage{lipsum}
\usepackage[T1]{fontenc}
\usepackage{array}
\usepackage{booktabs}
  
\usepackage{longtable,array,booktabs}
\usepackage{supertabular}
\usepackage{longtable}
\usepackage{graphicx}
\usepackage{amsmath}
\usepackage{amsfonts}
\usepackage{amssymb}
\usepackage{braket}
\usepackage{bm}
\usepackage{multirow}
\usepackage{color}
\usepackage[normalem]{ulem}
\usepackage{mathrsfs}
\usepackage{mathtools}
\usepackage{float}
\usepackage{graphicx}
\usepackage{amsfonts}
\usepackage{epsfig}
\usepackage{dcolumn}
\usepackage{amsthm}
\usepackage{color}

\usepackage{latexsym}
\usepackage{amscd}
\usepackage{hyperref}
\usepackage[mathscr]{eucal}
\usepackage{physics}

\usepackage{bm}      
\usepackage{ifpdf}
\usepackage{epstopdf}

\makeatletter

\newcommand{\Rmnum}[1]{\expandafter\@slowromancap\romannumeral #1@}
\makeatother

\hypersetup{colorlinks=true, linkcolor=blue, citecolor=blue, filecolor=blue, urlcolor=blue}
\AtBeginDocument{%
    \newwrite\bibnotes
    \def\bibnotesext{Notes.bib}
    \immediate\openout\bibnotes=\jobname\bibnotesext
    \immediate\write\bibnotes{@CONTROL{REVTEX41Control}}
    \immediate\write\bibnotes{@CONTROL{%
    apsrev41Control,author="08",editor="1",pages="1",title="0",year="1"}}
     \if@filesw
     \immediate\write\@auxout{\string\citation{apsrev41Control}}%
    \fi
}%

\begin{document}

%\preprint{APS/123-QED}

\title{Electron and phonon topology in transition metal material TaSi}% Force line breaks with \\

\author{Saurabh Kumar Sen}
\email{saurabhgsen@gmail.com}%
\author{Shivendra Kumar Gupta}%
\author{Nagarjuna Patra}%
\author{Ajit Singh Jhala}%
\author{Poorva Singh}%
 \email{poorvasingh@phy.vnit.ac.in}
\affiliation{
 Department of Physics, \\ Visvesvaraya National Institute of Technology, Nagpur, 440010, India \\
}%

\date{\today}% It is always \today, today,

\begin{abstract}

The plethora of multifold quasiparticles in topological materials has led to significant advancements in condensed matter physics, inspiring the investigation for materials that host both electronic and bosonic multifold excitations. In this work, we  explore the electronic and phononic properties of TaSi, a non-symmorphic chiral topological material crystallizing in space group $P2_1 3$ (No. 198). This system exhibits multifold fermions, which are higher-spin generalizations of Weyl fermions, protected by the unique crystalline symmetries of the structure. Using first-principles calculations, we predict that electronic band possesess fourfold spin-$\frac{3}{2}$
 Rarita-Schwinger (RSW) fermions, sixfold excitations (double spin-1), all possessing large Chern numbers $C = + 4$ and Weyl fermions of spin-$\frac{1}{2}$
 with Chern no. $-1$ in the presence of spin orbit coupling (SOC). Additionally, the phononic band structure hosts chiral bosonic excitations characterized by Chern numbers $C = \pm 2$. The coexistence of chiral electronic and bosonic quasiparticles give rise to exotic transport phenomena, rendering the material as promising candidate for future applications in quantum materials, topological electronics, and spintronics.
 
\end{abstract}

\maketitle

\section{\label{sec:level1}Introduction\protect}
In recent years, topological insulators (TIs) have attracted a great deal of attention for their exotic Dirac like surface states \cite{gupta2023coexistence, hasan2010colloquium} and has sparked a significant research interest in semimetals, including Weyl semimetals (WSMs)\cite{hirschberger2016chiral,huang2015theoretical,lv2015observation, shekhar2018anomalous,wan2011topological,xu2015discovery,weng2015weyl,weng2015weyl}, Dirac semimetals (DSMs) \cite{liu2014stable,liu2014discovery,xiong2015evidence,wang2012dirac} , nodal-line semimetals (NDLS) \cite{huang2016topological,fang2016topological}, and multifold fermionic system \cite{chang2017unconventional,tang2017multiple,rao2019observation,sanchez2019topological,takane2019observation, schroter2020observation}. Dirac and Weyl fermions, predicted in 1928 \cite{dirac1981principles} and 1929 \cite{weyl1929gravitation} respectively, have later been theoretically and experimentally identified in both two-dimensional (2D) and three-dimensional (3D) systems, where they have protected band crossing near the Fermi level. 

In Dirac semimetals, these bands crossing point, known as Dirac cone, is a fourfold degenerate due to the presence of time-reversal symmetry (TRS) and inversion symmetry (IS), which ensures double degeneracy of all bands by Kramers' theorem \cite{young2012dirac}. In contrast, WSMs host a pair of twofold-degenerate Weyl nodes, which arise when either TRS or IS, or both are broken. The separation of these nodes in momentum space determines the material’s topological strength. 

Each Weyl node  represents a singularity of the Berry curvature, effectively acting as a magnetic monopole in momentum space. These nodes occur in pairs with opposite chirality and serve as a topological source or sink of the Berry curvature associated with Bloch wavefunction\cite{ma2021observation}. The Berry curvature  arises from the topological entanglement between conduction and valence bands, playing a crucial role in defining the nontrivial topological character of Weyl semimetals.\cite{zhang2018berry}.

WSMs exhibit unique properties such as open Fermi arcs in their surface electronic band structure, the chiral anomaly and negative magnetoresistance. The open Fermi arc offers strong evidence for identifying a WSM, as observed through angle-resolved photoemission spectroscopy (ARPES)\cite{sanchez2018discovery,schroter2019chiral,rao2019new}. These materials have potential for novel superconductivity and next-generation spintronics or electronics devices.

The Hamiltonian governing a three-dimensional massless spin-\(\frac{1}{2}\) Weyl fermion, as determined by the spin-statistics theorem, is expressed as $H = \eta \hbar v \mathbf{k} \cdot \boldsymbol{\sigma}$ \cite{barman2020symmetry} where $\eta$ represents the chirality of Weyl fermions. This expression is characterized by its linear dependence on momentum \(\mathbf{k}\) and the presence of a characteristic velocity \(v\). The Pauli matrices \(\boldsymbol{\sigma}\) represent the spin-\(\frac{1}{2}\) nature of the fermion, describing its intrinsic degree of freedom.

Beyond Dirac and Weyl fermions, diverse quasiparticles in condensed matter, distinct from those in high-energy physics, have gained signiﬁcant attention.  Lev Landau's theory of Fermi-liquid in 1930 provided insight into low-energy excitation, showing that they could be treated as nearly independent particle. This led to the concept of quasiparticles- effective entities that behave like free particles but incorporate interaction effects \cite{landau1957theory}. In 2016, Bradlyn \textit{et al.} \cite{bradlyn2016beyond} and Wieder \textit{et al.} \cite{wieder2016double} demonstrated that non-symmorphic crystalline symmetries can give rise to highly degenerate band crossings, which are absent in high-energy physics framework. This discovery emphasized the significance of condensed matter physics as a fertile ground for uncovering new fermionic particles and quantum phenomena. Unlike the strict requirements of full Poincaré symmetry, Lorentz invariance or the spin-statistics connection in  high-energy physics, condensed matter physics offers more flexibility \cite{tang2019comprehensive,zhang2019catalogue}. 

These low-energy quasiparticles are protected by chiral crystal symmetry and host three, four, or sixfold band degeneracy known as multifold fermions  \cite{bradlyn2016beyond}. The generalised Hamiltonian is described as  $H = \eta \hbar v \mathbf{k} \cdot \boldsymbol{S}$ \cite{barman2020symmetry}  where S represents an effective spin degree of freedom, which can take values corresponding to both integer and half-integer spin.

%Material with multifold fermions can also host Weyl fermions, which may exhibitsspecial properiteis in chiral structure such as quantized circular photogalvanic effect (CPGE), gyrotropic magnetic effect, etc

%%%%%%%%%%%%%%%%%%%%%%%
\begin{figure}[t]
\centering
\includegraphics[width=0.8\linewidth]{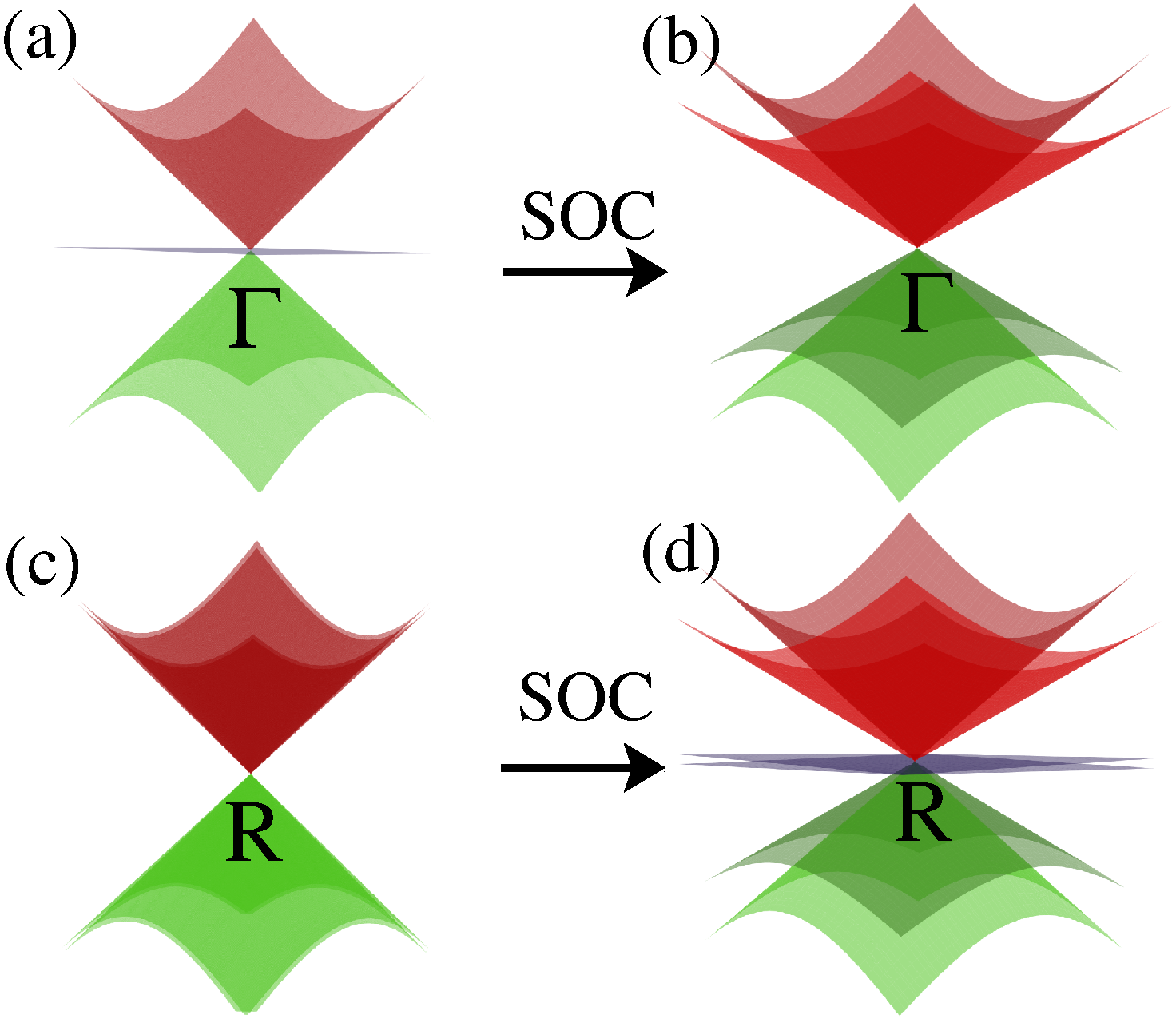}% Here is how to import EPS art
\caption{\label{fig:schematic.png} Classification of topological fermions based on band crossing and spin characteristics: (a) three-fold at $\Gamma$ with spin-1. (b) fourfold  RSW fermion at $\Gamma$ with spin-$\frac{3}{2}$. (c) fourfold at $R$  with double spin-$\frac{1}{2}$. (d) sixfold at $R$ with  double spin-1} 
\end{figure}

%%%%%%%%%%%%%%%%%%%%%%%%%%%%%%%%%%%%%%%%%%
Fermionic system has extensively demonstrated the role of topological properties in shaping diverse physical phenomena in material science, similarly bosonic quasiparticles such as phonons, which represent quantized lattice vibrations are increasingly being recognized as promising platform for exploring analogous topological effect \cite{li2012colloquium}.
 
Classical acoustic phonons, which correspond to low-energy sound waves in macroscopic system, are typically limited to kHz frequencies. These low-energy phonons do not significantly influence fundamental physical process such as heat conduction or electron-phonon coupling. Instead, such process are primarily governed by high-energy THz phonons, which arise from atomic-scale lattice vibrations in solid \cite{li2021computation}. Materials with nonsymmorphic symmetry  leading to the creation of unusual phonon behavior  are particularly interesting due to it's accessibility in  THz frequency range. Phonons in this frequency range play a key role in areas like  topological phononics \cite{liu2018berry}, phonon waveguides \cite{liu2017model} and advanced thermal management \cite{tang2021topological}.

Since the entire phonon spectrum is experimentally accessible, it provides an opportunity to study topological band crossing across the spectrum, which can be a significant advantage. Additionally, phonons are spinless which  make them unaffected by spin–orbit coupling (SOC)\cite{aiswarya2024insights}. As a result, topological phonons serve as an excellent platform for exploring spinless topological materials. 

Recent studies have proposed various materials, including transition metal sulfides and silicides, KMgBO$_3$ \cite{sreeparvathy2022coexistence} Zr$_3$Ni$_3$Sb$_4$\cite{zhong2021coexistence}, Li$_3$CuS$_2$ \cite{liu2021charge}, BiIrSe \cite{liu2021charge}, and MSi(M = Fe,Co,Mn,Re, or Ru)\cite{zhang2018double}, AlPt \cite{schroter2019chiral}, SiTc \cite{gupta2025topologicalelectronicphononicchiral,gupta2025topological}. These materials have special symmetries that allow the formation of Weyl-like states in their phononic system. Specifically, in FeSi material \cite{miao2018observation}, double Weyl phonons have been experimentally observed. 

In topological materials, the coexistence of fermionic (electronic) and bosonic (phononic)  within a single crystal material plays a crucial role in enhancing both thermoelectric performance \cite{singh2018topological} and superconductivity \cite{dong2022superconductivity}. The topological nature of these excitation can significantly reduce phonon-mediated thermal conductivity while maintaining high electrical conductivity, leading to an enhancement of the thermoelectric figure of merit (zT). Simultaneously, such materials can exhibit unconventional superconductivity, where the interplay between topological electron and phonon strengthen the pairing mechanism, further enriching their quantum transport properties.

In this work, based on the first-principle calculations, we conduct a systematic investigation of topological electronic and bosonic properties of TaSi material which offer new insight into its band structure and phonon dynamics. We find that the material shows  coexistence of four fold excitation with chiral spin-$\frac{3}{2}$ Rarita-Schwinger (RSW)
 fermions and sixfold excitation (double spin-1 fermions) as well as spin-$\frac{1}{2}$ Weyl points near the Fermi level. Our findings provide the compelling evidence of fermionic and bosonic quasiparticles in condensed matter system.
%%%%%%%%%%%%%%%%%%%%%%%%%%%%%%%%%%%%%%%%%%%%%
\begin{figure}[t]
\includegraphics[width=0.8\linewidth]{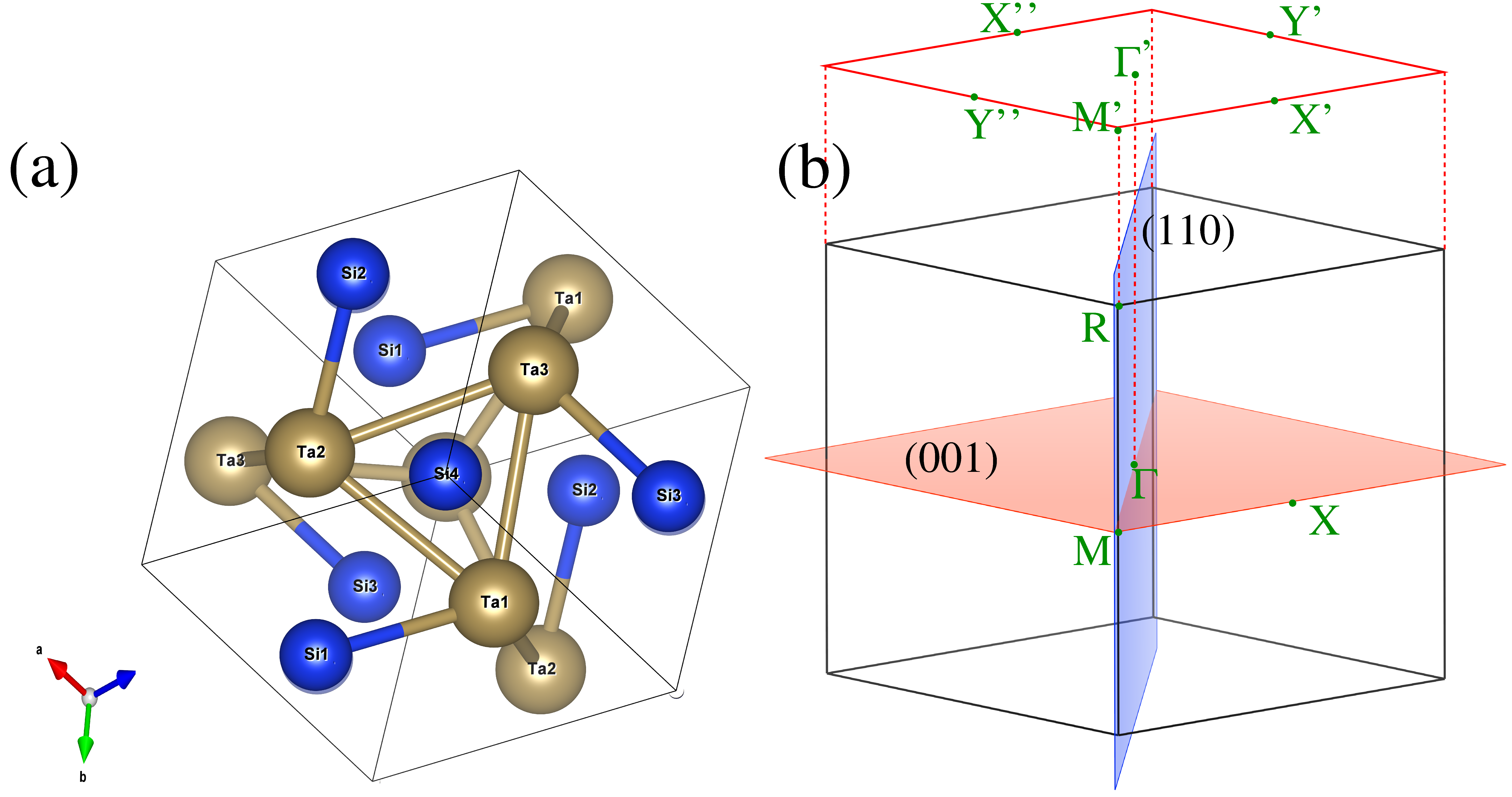}% Here is how to import EPS art
\caption{\label{fig:XSi.png} Crystal structure of TaSi  $P2_1 3$ (No. 198)  (a) Bulk crystal structure  (b)  Bulk and surface Brillouin zone along (001) and (110)  with different colors.}
\end{figure}
%%%%%%%%%%%%%%%%%%%%%%%%%%%%%%%%%%%%%%%%%%%%%%%%%%%%%%%%%%%%%%%%%%%%%%%%%%%%%%%%%%
\section{Result and Discussion}
The material TaSi belonging to a cubic crystal with space group P$2_1$3 (198) is a chiral crystal whose symmetry enforces it to host multifold fermions as low-energy quasiparticles. Fig.~\ref{fig:schematic.png}(a) and  ~\ref{fig:schematic.png}(b) show schematic representations of multifold fermions at the $\Gamma$ time revarsal invarient momenta (TRIM) point, while fig.~\ref{fig:schematic.png}(c) and (d) show the corresponding representations at the $R$ TRIM point,  without and with SOC, respectively. Fig.~\ref{fig:XSi.png} shows the three-dimensional (3D) crystal structure and it's bulk and surface Brillouin zone are shown along the (001) and (110) planes.

\section{Electronic Band Structure}
\begin{figure}[!t]
\centering
\includegraphics[width=\linewidth]{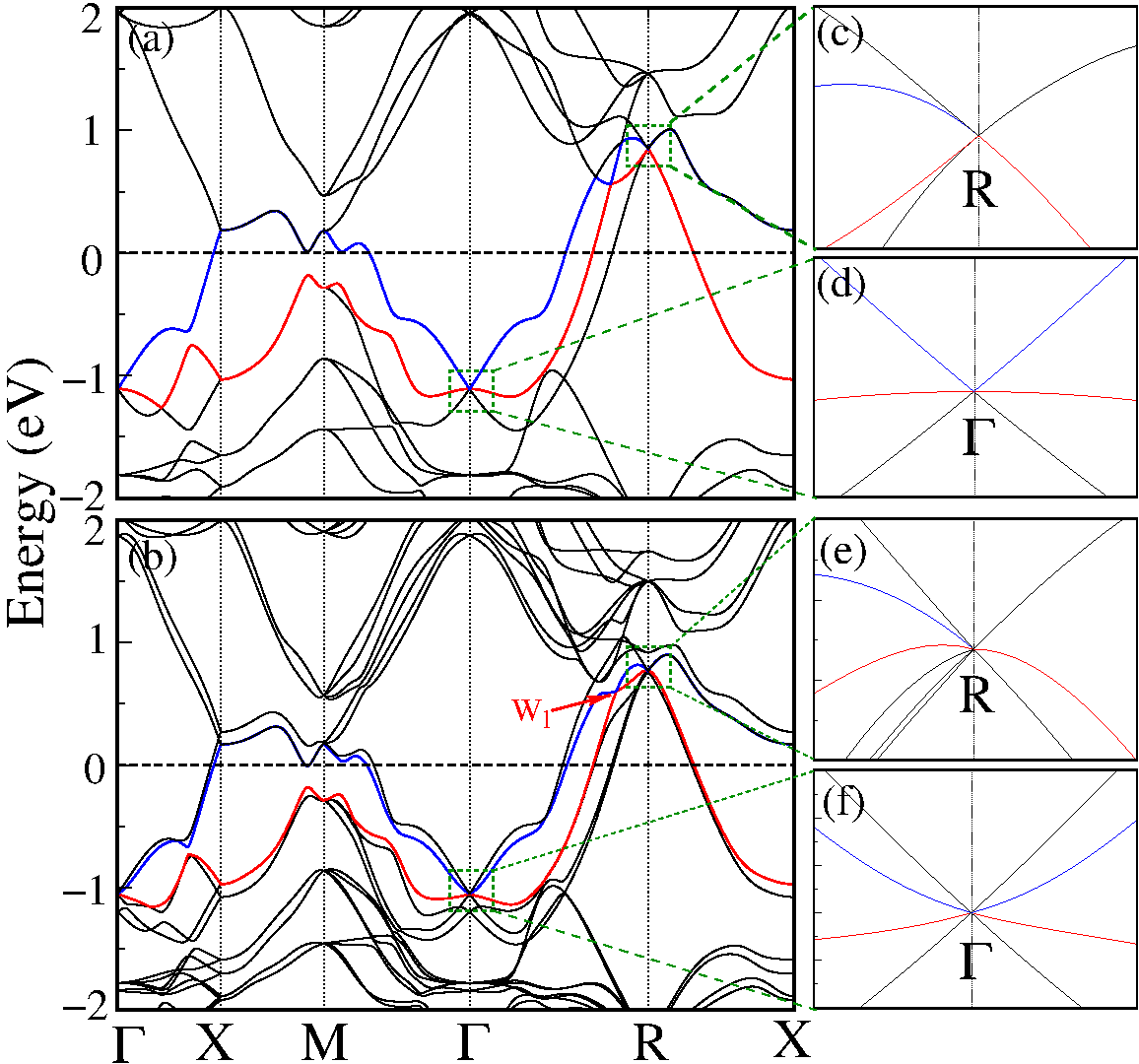}
\caption{\label{fig:band_Ta.png} The electronic bulk band structure along high-symmetry paths (a) without spin–orbit coupling (SOC), and (b) with SOC. (c) and (d) show the enlarged views at the $\Gamma$ and $R$ points without SOC, respectively, while (e) and (f) present the corresponding enlarged views with SOC at the $\Gamma$ and $R$ points, respectively.}
\end{figure}

The topological characteristics of a material are intrinsically reflected in its electronic band structure. In this section, we present the electronic band structure of TaSi  obtained via ab initio calculation, with a particular focus on the high-symmetry points $\Gamma$, $R$ and along $\Gamma$ to R high symmetry line in the Brillouin zone (BZ). Fig.~\ref{fig:band_Ta.png}  illustrate the band structure of TaSi. The band crossing occur at $-1.05$ eV at $\Gamma$, $0.61$ eV along $\Gamma$ to R  and $0.76$ eV at $R$, positioning them in proximity to the Fermi level.

\begin{figure}[h]
\centering
\includegraphics[width=1\linewidth]{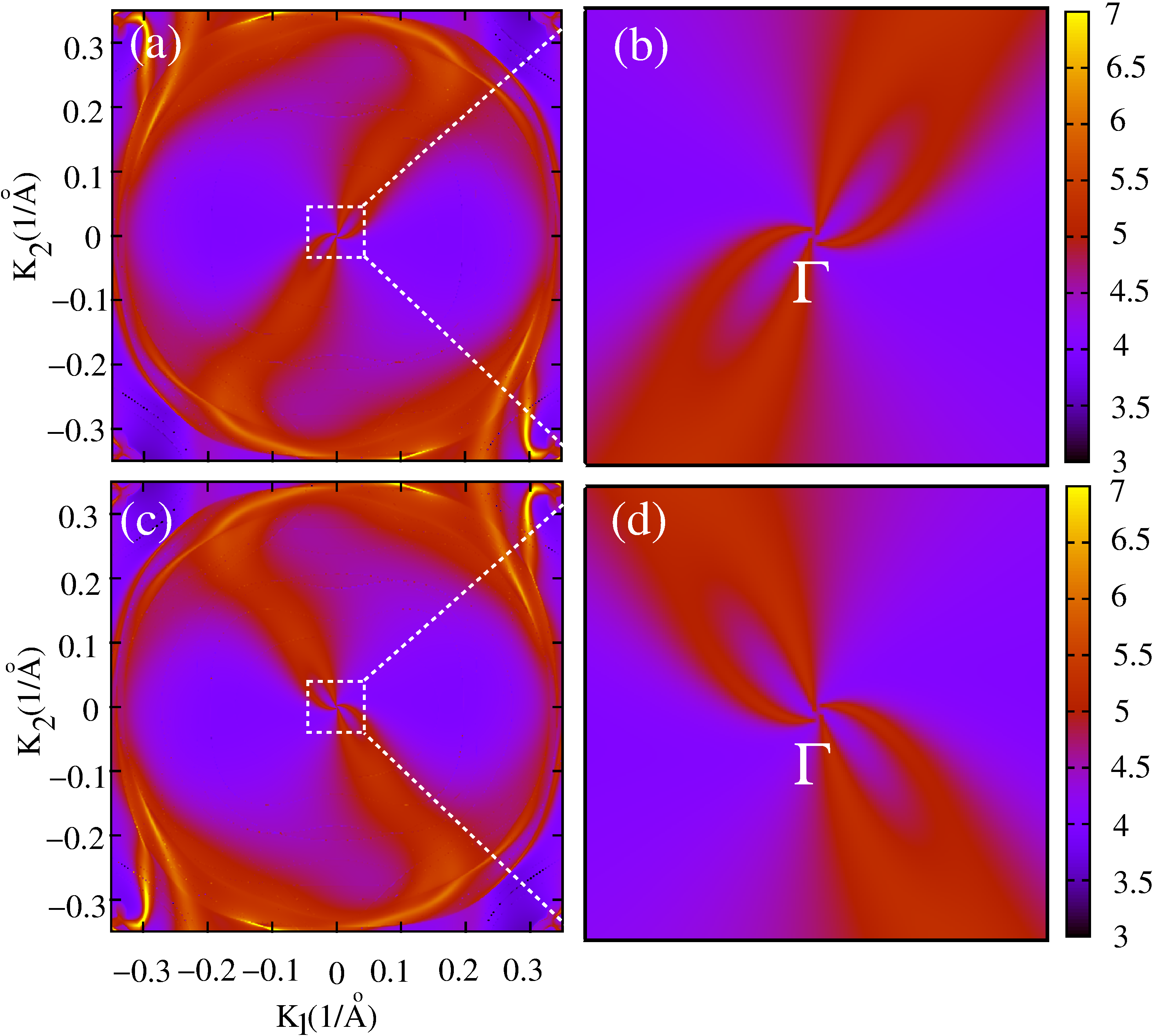}% Here is how to import EPS art
\caption{\label{fig:arc_TaSi.png} Fermi arcs at the $\Gamma$ point along (001) direction, (a)–(c) show the chiral Fermi arcs associated with a chirality of $+4$, (b) and (d) present the enlarged views, highlighting the emergence of four Fermi arcs associated with spin-$\frac{3}{2}$ RSW fermion.}
\end{figure}

Fig.~\ref{fig:band_Ta.png}(a) depicts the band structure of the material in the absence of spin-orbit coupling (SOC), where characteristic threefold and fourfold band crossing is observed at the $\Gamma$ and $R$ points, corresponding to spin-1 and double spin-$\frac{1}{2}$
 fermions, respectively. At the $\Gamma$, the threefold degeneracy is protected by two screw symmetries, $S_{2z}$, $S_{2y}$, and a threefold rotation $S_3$, which satisfy the condition that the two screw symmetries $S_{2y}$ and $S_{2z}$ must commute and square to the identity, i.e., $S_{2y}^2$ $=$ $S_{2z}^2 = I$. Additionally, the threefold rotation $S_3$ should act nontrivially, meaning $S_3 \ket{\psi} \neq \ket{\psi}$, where $\ket{\psi}$ is a simultaneous eigenstate of $S_{2y}$ and $S_{2z}$  that leads to threefold degeneracy. Meanwhile, at the $R$-point, screw symmetries $S_{2x}$ and $S_{2y}$, which anticommute ($\{S_{2x}, S_{2y}\} = 0$) and square to $-I$, combined with a three-fold rotation $S_3$, ensures the formation of a symmetry-protected fourfold degeneracy, as the screw symmetries generate orthogonal states that cannot be further split in a spinless system \cite{barman2020symmetry, bradlyn2016beyond}.

To highlight these  features, the band structures near these TRIM points are enlarged in fig.~\ref{fig:band_Ta.png}(c) and~\ref{fig:band_Ta.png}(d). As shown in fig.~\ref{fig:band_Ta.png}(c), four bands intersect at the $R$ point, whereas three bands cross at the $\Gamma$ point in fig.~\ref{fig:band_Ta.png}(d). Also  there is one crossing W1 along the high symmetry line $\Gamma$ to R, which corresponds to spin-$\frac{1}{2}$ Weyl fermion.

With the inclusion of SOC,  time-reversal symmetry satisfies $T^2 = -I$, that causes the degeneracy of electronic band structure, a  fourfold  at the $\Gamma$  and a sixfold at the $R$ TRIM point as shown in fig.~\ref{fig:band_Ta.png}(b) \cite{barman2020symmetry, bradlyn2016beyond}, and  one Weyl point exist along $\Gamma$ to $R$ line  with spin-$\frac{1}{2}$. At $\Gamma$, the fourfold degeneracy evolves  because of the anticommuting screw symmetries that square to $-I$ as evident from fig.\ref{fig:band_Ta.png}(f). In contrast, at the $R$ TRIM point, Kramers degeneracy applies, and sixfold degeneracy  is protected, distinguishing $R$ from $\Gamma$, as shown in fig.~\ref{fig:band_Ta.png} (e). Thus, $\Gamma$ supports at most fourfold degeneracy due to its higher symmetry constraints and $R$ allows six fold degeneracy, highlighting how SOC and symmetry interplay shapes the multifold fermionic excitations in TaSi material.

\begin{figure}[h]
\includegraphics[width=1.0\linewidth]{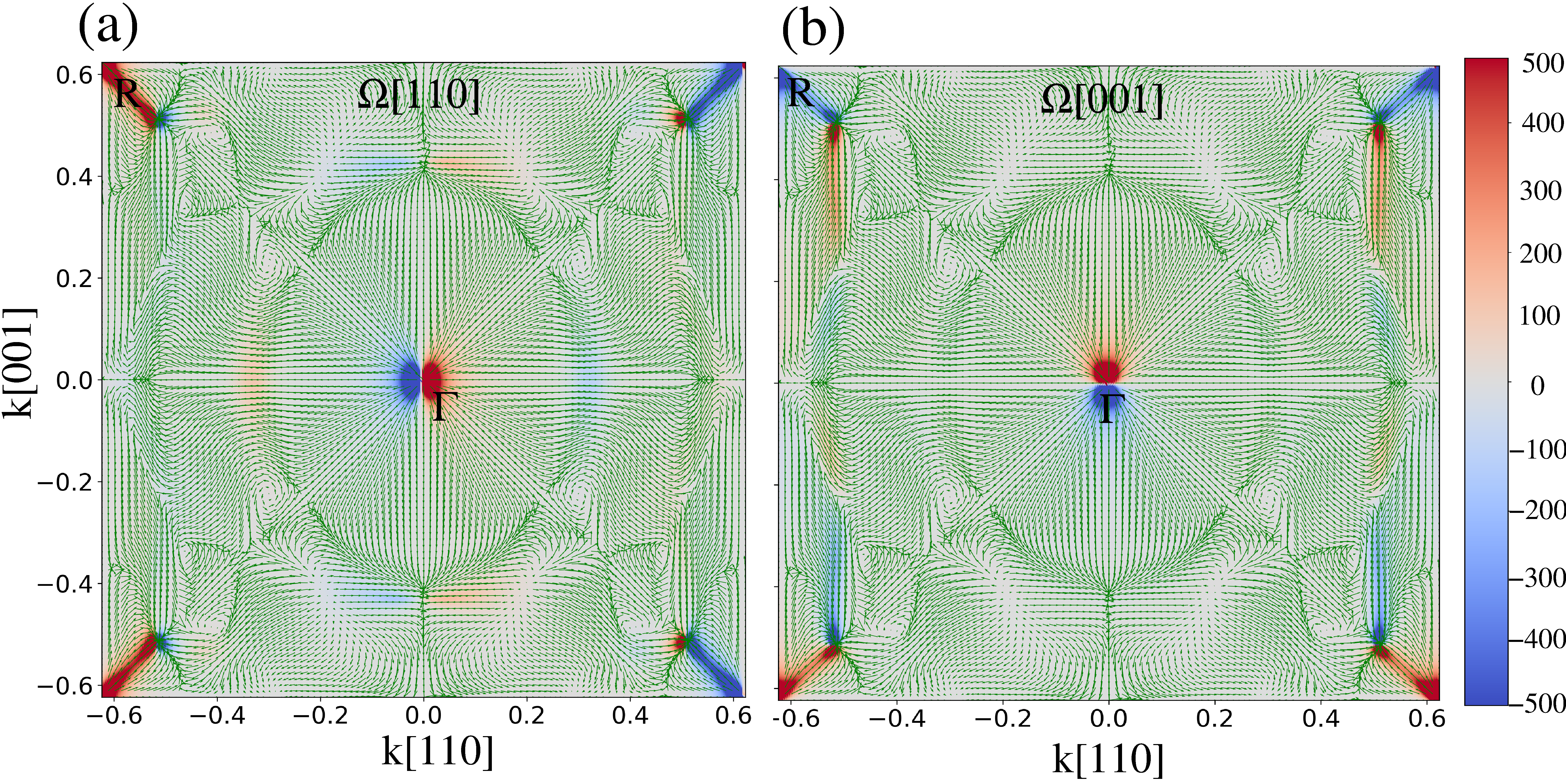}% Here is how to import EPS art
\caption{\label{fig:Berry_electronic.png} Berry curvature distribution plotted in the (110) plane for (a) TaSi. The plots clearly reveal the locations of the multifold fermions and W1 Weyl fermion, characterized by their respective spin and chirality: $+4$, $+4$, and $-1$. These features are consistent with the sign and magnitude of the topological charges in the presence of SOC}
\end{figure}
 
In the first BZ, two multifold fermions and eight Weyl points have been identified, which can be distinguished based on their chirality. Specifically, eight Weyl points of the W1 type exhibit a chirality of -1 each, contributing to a total chirality of -8. In contrast, a fourfold fermion at $\Gamma$ point and a six fold fermion at R point both possess a chirality of +4, collectively contributing +8 to the total chirality. The summation of these contribution, results in an overall chirality of zero, satisfying the topological WSM requirement dictated by the Nielsen-Ninomiya theorem \cite{PhysRevLett.20.695, nielsen1981no, bruni2001gravitational}. 

To further confirm the topological characteristics, we carry out a Fermi arc analysis, presented in fig.~\ref{fig:arc_TaSi.png}, which display the two-dimensional projections along the (001) direction. Fig.~\ref{fig:arc_TaSi.png}(a) confirms the presence of four Fermi arcs emerging from the $\Gamma$ TRIM point. These Fermi arcs are consistent with the total chirality of $+4$, which is characteristic of spin-$\frac{3}{2}$ fermions.
The chiral characteristics of these Fermi arc is further illustrated in fig.~\ref{fig:arc_TaSi.png}(c). For enhanced visibility, the Fermi arcs have also been magnified in separate panels, fig.~\ref{fig:arc_TaSi.png}(b) and fig.~\ref{fig:arc_TaSi.png}(d). 

In addition, we have also verified the electronic topolgy in TaSi through the Berry curveture distribution, as shown in fig.~\ref{fig:Berry_electronic.png}. The results indicate a monopole-like source of Berry curvature with a net chirality of $+4$ located at both the \( \Gamma \) and \( R \) high-symmetry points, corresponding to the presence of a spin-$\frac{3}{2}$ and double spin 1 
 fermions, respectively. These chiralties are compensated by eight sinks with chirality \( -1 \) distributed along the \( \Gamma \)-\( R \) high-symmetry line. The net chiral charge is fully compensated by opposite chirality nodes, confirming the topological Weyl semimetal nature of the material.

\section{Phonon}
\begin{figure}[t]
\includegraphics[width=1.0\linewidth]{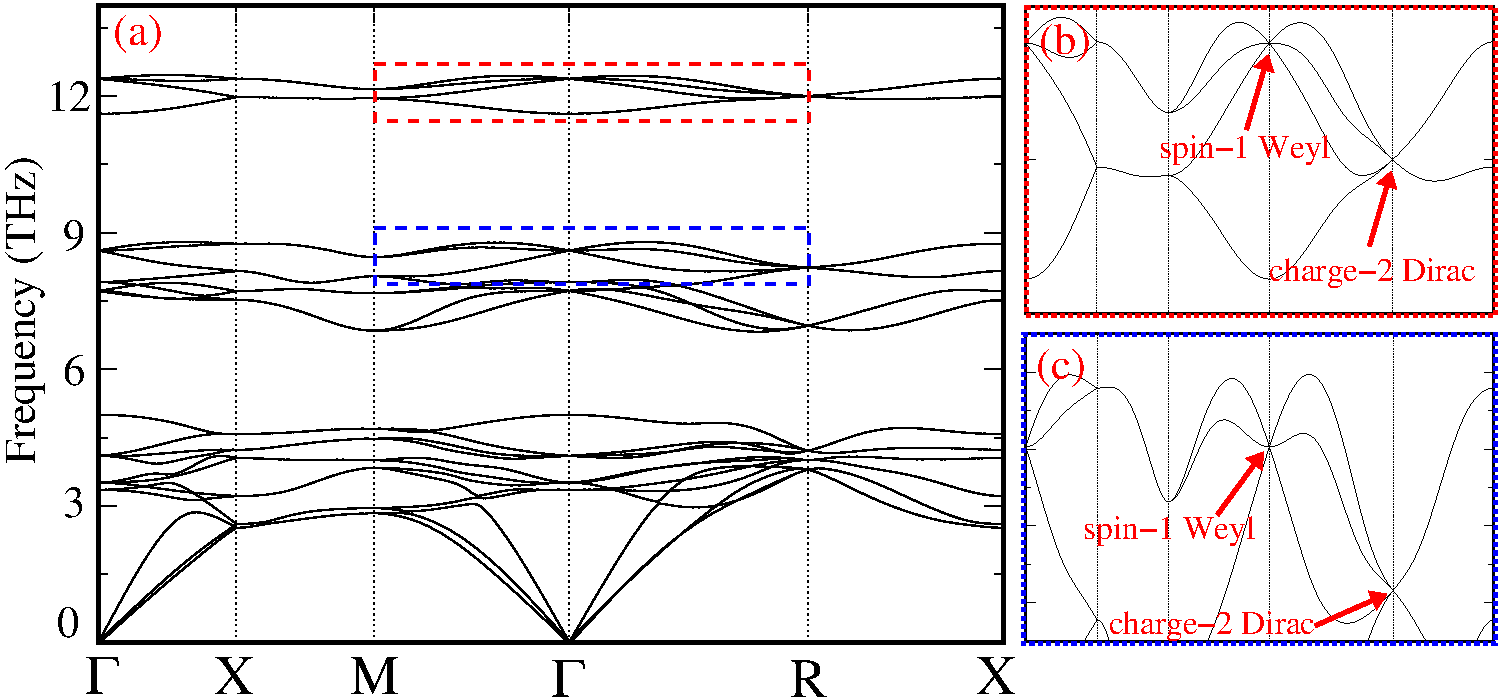}% Here is how to import EPS art
\caption{\label{fig:phon.png} Phonon dispersion curve showon in (a),  The insets (b) and (c) provide enlarged views around the $\Gamma$ point, highlighting the spin-1 Weyl fermions, and around the $R$ point, indicating the charge-2 Dirac features in the high-frequency region.}\end{figure}

The material TaSi consisting of 8 atoms (four of Ta and four of Si) shows 24 phonon modes out of which 3 are associated with acoustical modes and rest are optical modes. Fig.~\ref{fig:phon.png} present the computed phonon dispersion curve for TaSi. The absence of imaginary frequencies confirms the dynamical stability of the material, indicating that it remains structurally stable under perturbations. 

In fig.~\ref{fig:phon.png}(a), we focus on the phonon band characteristics near the 8-9 and 11.5-12.5 THz frequency ranges, as these frequency ranges significantly influence topological properties of the material. The enlarged view of these specific frequency ranges are presented in fig.\ref{fig:phon.png}(b) and fig.\ref{fig:phon.png}(c), where notably, a spin-1 Weyl point characterized by a threefold band crossing emerges at the $\Gamma$ high-symmetry point, while a charge-2 Dirac point associated with a fourfold band crossing observed at the R high-symmetry point. These features confirmed the presence of symmetry-protected crossings, which are essential for identifying the phononic topological nature of the chiral crystal system.

\begin{figure}[t]
\includegraphics[width=1.0\linewidth]{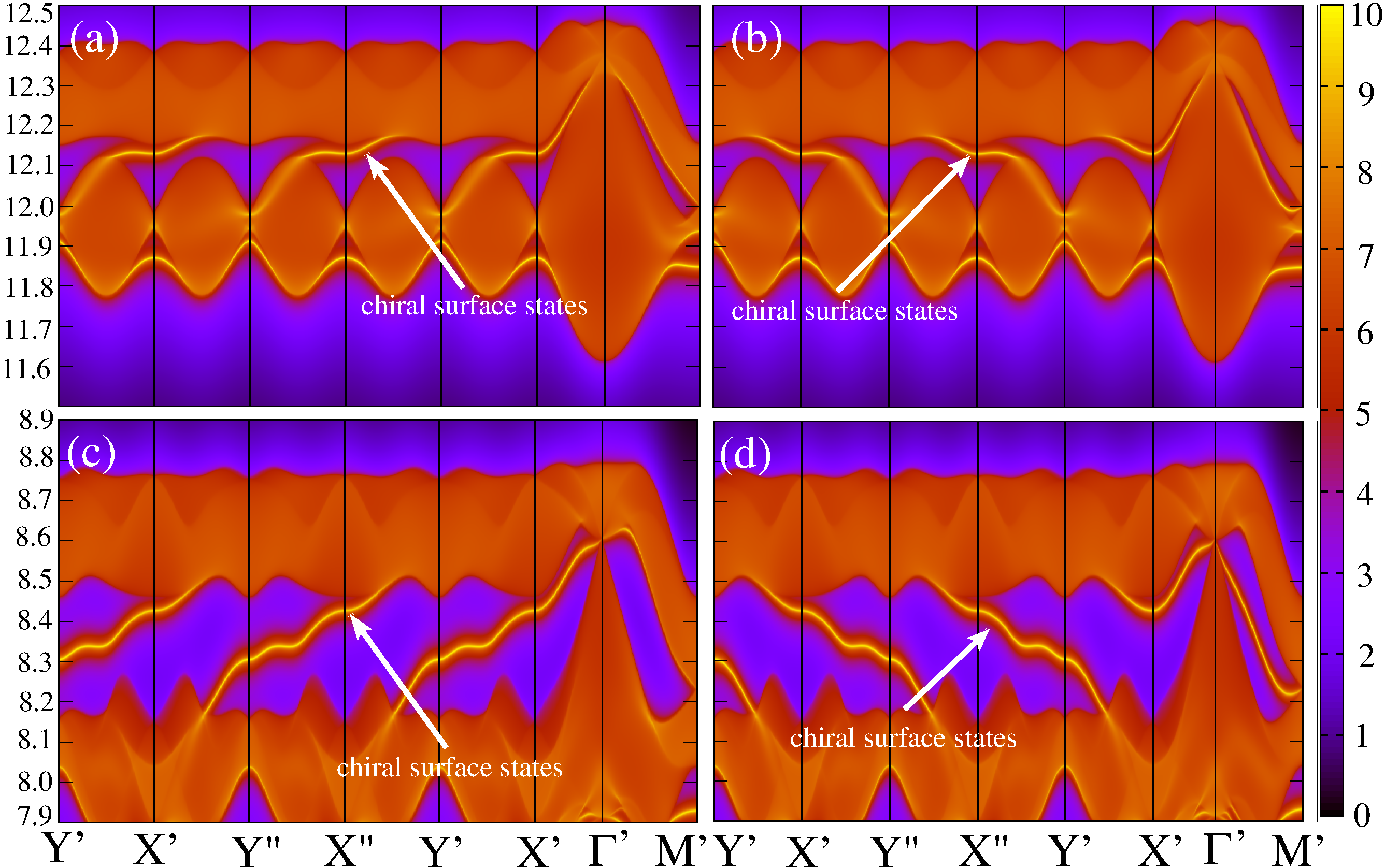}% Here is how to import EPS art
\caption{\label{fig:full_chiral.png} Phononic surface states, (a)-(b) correspond to the frequency range 11.5-12.5 THz (c)-(d) depict the states in the 8-9 THz range.}
\end{figure}
 
\begin{figure}[t]
\includegraphics[width=1.0\linewidth]{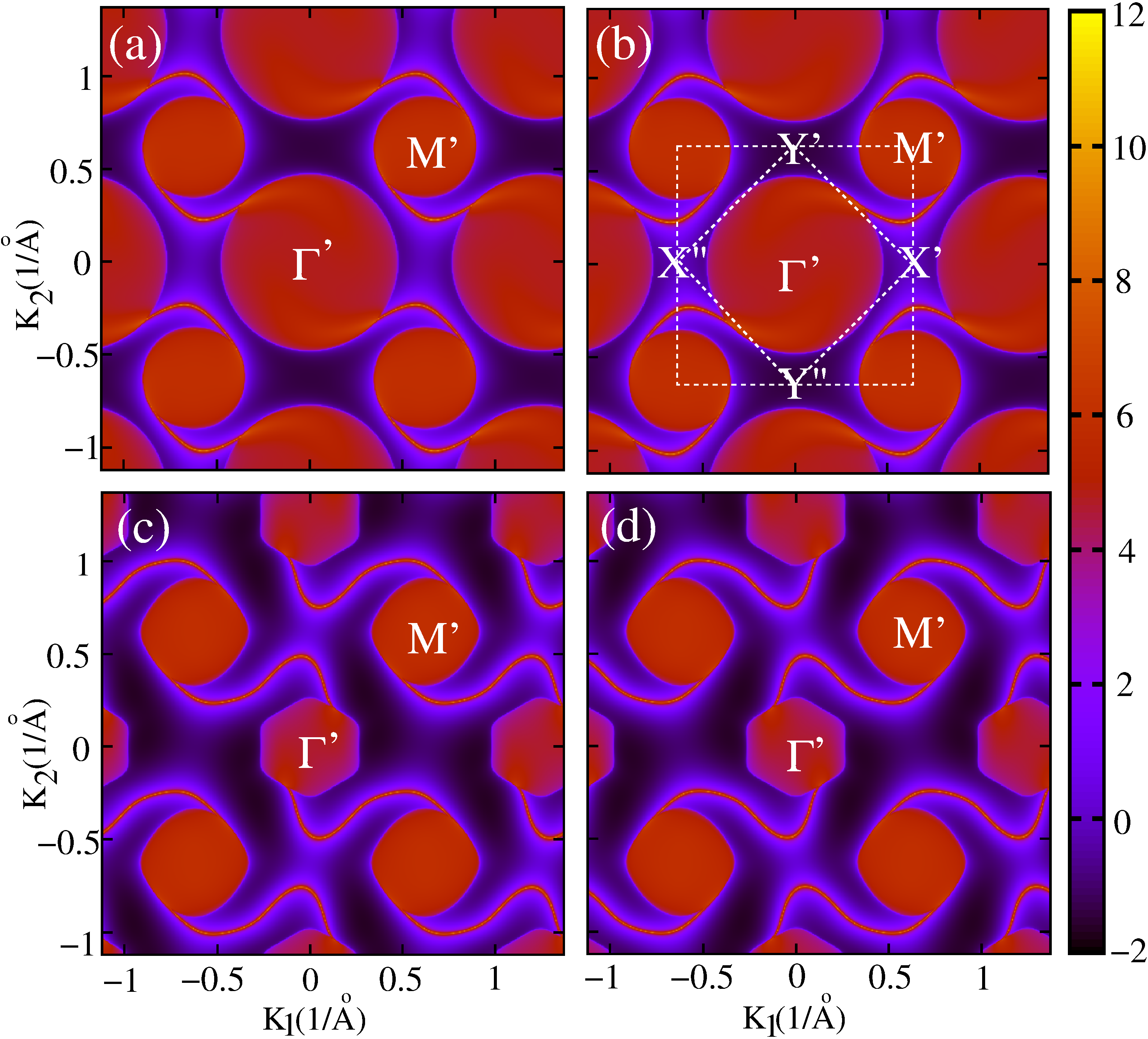}% Here is how to import EPS art
\caption{\label{fig:full_arc.png} Fermi arcs  projected along the (001) surface. (a)-(b) correspond to  11.5-12.5 THz, while (c)-(d) show  8-9 THz frequency range.}
\end{figure} 

\begin{figure}[t]
\includegraphics[width=1.0\linewidth]{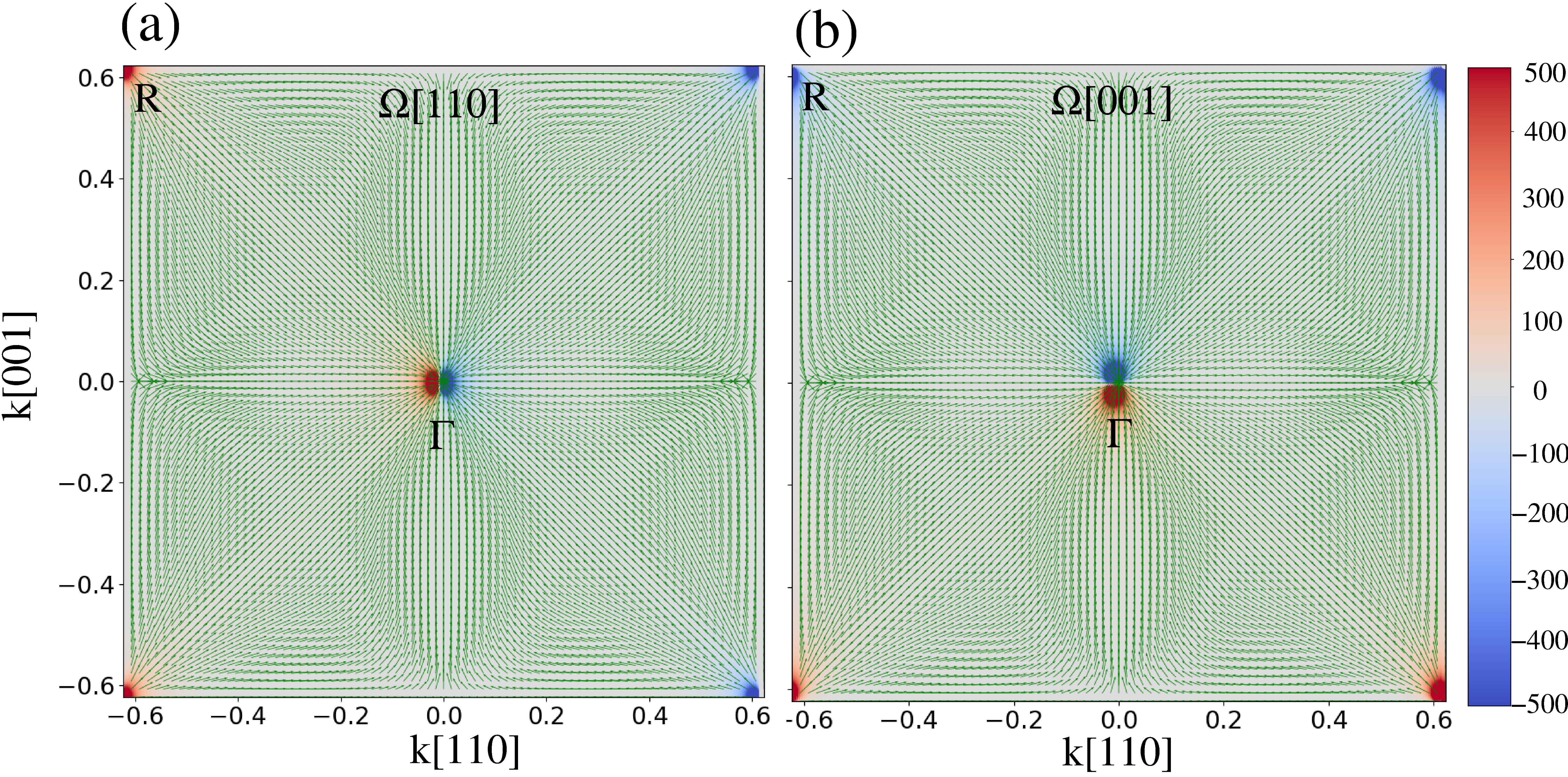}% Here is how to import EPS art
\caption{\label{fig:PhBerry_electronic.png} Berry curvature distribution (a) $\Omega$ [110] and (b) $\Omega$ [001].}
\label{surfaceTaV}
\end{figure}

To further confirm the presence of chiral states in TaSi, we have computed the phononic surface states for (001) surface, as presented in fig.~\ref{fig:full_chiral.png}. The analysis is performed along the path Y'-X'-Y"-X"-Y'-X'-$\Gamma$'-M'. Here, $\Gamma$' and M' represent the projections of bulk TRIM points $\Gamma$ and R, respectively (as shown in fig~\ref{fig:XSi.png}). It is evident that the surface states emerging from the $\Gamma$' point extend towards the M' high-symmetry point- a feature consistently observed in both selected frequency ranges.

Moreover the surface states also allow for identification of the Fermi arcs formed by the chiral states between $\Gamma$' and M' high symmetry points as shown in fig.\ref{fig:full_arc.png}. Since $\Gamma$' and M' are connected by  surface states, the presence of oppositely charged Weyl points are expected, leading to the emergence of a Fermi arc.

Fig.~\ref{fig:full_arc.png} (a)-(b),  depicts chiral counterparts of the Fermi arc, corresponding to the 11.5–12.5 THz range, while fig.\ref{fig:full_arc.png}(c)–(d) represents the arcs assosciated  with the 8-9 THz range  projected onto the (001) surface. It is evident from the fig.\ref{fig:full_arc.png}(a)-(d) that there are two arcs emerging from the $\Gamma$ and connecting to M' high symmetry points agrring with the chirality $\pm2$.

We validate the chiral phononic topology in TaSi through Berry curvature distribution, with the vector field $\Omega$ evaluated on selected momentum space planes, as shown in fig.\ref{fig:PhBerry_electronic.png}. On the $k_x$-$k_y$ plane, the vector field of $\Omega$ exhibits direct flow from the $\Gamma$ to the $R$ point, with minimal out-of-plane deviation, supported by the intensity distribution of the $xy$ [fig.\ref{fig:PhBerry_electronic.png}(a)] and $z$-components [fig.\ref{fig:PhBerry_electronic.png}(b)] of the Berry curvature. Notably, the $\Gamma$ and $R$ points act as local source and sink of Berry flux respectively,  with Chern numbers $C = \pm2$, signify the presence of unconventional chiral fermions and highlight the nontrivial topology of the bulk bands.

\section{Computational Methods}
Our first-principles calculations were performed within the framework of density functional theory (DFT) \cite{kohn1965self}, using a plane-wave basis set and the projector augmented wave (PAW) method \citep{torrent2010electronic}, as employed in the Vienna Ab initio Simulation Package (VASP) \cite{hafner2008ab, blochl1994projector, kresse1996efficient}. The generalized gradient approximation (GGA) with the Perdew-Burke-Ernzerhof (PBE) functional was used to describe the exchange-correlation interactions \cite{perdew1996generalized}.
The plane-wave energy cutoff was set to 1.5 times the default cut off energy(500 eV) and a 12 × 12 × 12 Monkhorst-Pack kmesh was used for the BZ sampling. The structure was fully optimized, with the Hellmann-Feynman forces converged to a tolerance of $10^{-4} \, \text{eV/Å}$. The convergence criterion for the total energy was set to be $10^{-8} \, \text{eV}$. To calculate the topological features, a tight-binding Hamiltonian was constructed by maximally localized Wannier functions implemented in 
\textnormal{W\scalebox{0.80}{\textsc{ANNIER90}}}  package \cite{marzari1997maximally,marzari2012maximally} and the iterative Green’s function method with \textnormal{W\scalebox{0.80}{\textsc{ANNIERTOOLS}}} \cite{wu2018wanniertools} package .

The phonon calculations were determined using Density Functional Perturbation Theory (DFPT), implemented in the phonopy package \cite{togo2015first}. We construct the supercell with lattice constant larger than 10~\text{\AA}  for TaSi material with dimensions of  $3\times 3\times 3$  to calculate the phonon band structure. Topological phonon surface state and arc were calculated using \textnormal{P\scalebox{0.80}{\textsc{HONOPYTB}}} \cite{phonopyTB} tool assist in \textnormal{W\scalebox{0.80}{\textsc{ANNIERTOOLS}}} \cite{wu2018wanniertools}.

\section{Conclusion}
In this study, we systematically investigate TaSi material, which crystallizes in the noncentrosymmetric space group P$2_1$3 and host exotic multifold fermions at their high-symmetry points in both electronic and bosonic (phononic) spectra. The topologically nontrivial nature of the material is established through comprehensive calculations, including electronic band structure analysis, surface state and Fermi arc analysis. TaSi emerges as a novel class of material exhibiting both bosonic and electronic topological properties. Berry curvature distribution confirm the chirality +4 at $\Gamma$, R and -1 along $\Gamma$ to R line  for electronic states and ±2 chirality for bosonic (phononic) states at the $\Gamma$ and $R$ TRIM points. This work highlights the coexistence of multiple topological electronic and phononic excitations, positioning TaSi material as promising candidate for future  experimental investigations in topological quantum materials.

\section{Acknowledgments} 
This study received financial support from the Anusandhan National Research Foundation (ANRF), Government of India, under Grant No. CRG/2022/006419, and from the Council of Scientific and Industrial Research -Human Resource Development Group (CSIR-HRDG) via the ASPIRE Grant No. 03WS(006)/2023-24/EMR-II/ASPIRE. The authors also acknowledge financial assistance from the Department of Science and Technology (DST) under the DST-FIST Project No. SR/FST/PSI/2017/5(C), to the Department of Physics at VNIT Nagpur. The authors express their sincere thanks to VNIT Nagpur for providing the computational facilities and infrastructure essential for this work. PS acknowledges the High-Performance Computing (HPC) support from NPSF C-DAC Pune. SKS is also grateful to the Ministry of Human Resource Development (MHRD), Government of India, for the institute fellowship.

\bibliography{bib}
\end{document}